\documentclass[fleqn,usenatbib]{mnras}

\usepackage{newtxtext,newtxmath}

\usepackage[T1]{fontenc}

\DeclareRobustCommand{\VAN}[3]{#2}
\let\VANthebibliography\thebibliography
\def\thebibliography{\DeclareRobustCommand{\VAN}[3]{##3}\VANthebibliography}

\usepackage{graphicx}	
\usepackage{amsmath}	
\usepackage{bbding}	
\usepackage{color}
\usepackage{gensymb}
\usepackage{comment}
\usepackage{empheq}

\usepackage{scalerel,tikz}
\usetikzlibrary{svg.path}
\definecolor{orcidlogocol}{HTML}{A6CE39}
\tikzset{orcidlogo/.pic={
 \fill[orcidlogocol] svg{M256,128c0,70.7-57.3,128-128,128C57.3,256,0,198.7,0,128C0,57.3,57.3,0,128,0C198.7,0,256,57.3,256,128z};
 \fill[white] svg{M86.3,186.2H70.9V79.1h15.4v48.4V186.2z}
 svg{M108.9,79.1h41.6c39.6,0,57,28.3,57,53.6c0,27.5-21.5,53.6-56.8,53.6h-41.8V79.1z M124.3,172.4h24.5c34.9,0,42.9-26.5,42.9-39.7c0-21.5-13.7-39.7-43.7-39.7h-23.7V172.4z}
 svg{M88.7,56.8c0,5.5-4.5,10.1-10.1,10.1c-5.6,0-10.1-4.6-10.1-10.1c0-5.6,4.5-10.1,10.1-10.1C84.2,46.7,88.7,51.3,88.7,56.8z};
}}
\newcommand\orcidicon[1]{\href{https://orcid.org/#1}{\mbox{\scalerel*{
\begin{tikzpicture}[yscale=-1,transform shape]
\pic{orcidlogo};
\end{tikzpicture}
}{|}}}}

\newcommand{\aref}[1]{\hyperref[#1]{Appendix~\ref{#1}}}

\defcitealias{2018MNRAS.477.2716K}{KBFC18}

\usepackage[normalem]{ulem}

\definecolor{darkgreen}{rgb}{0.13, 0.55, 0.13}
\definecolor{brown}{rgb}{0.65, 0.16, 0.16}

\title[RMHD PopIII simulations]{Magnetic fields limit the mass of Population III stars even before the onset of protostellar radiation feedback}

\author[Sharda et al.]{Piyush Sharda$^{\orcidicon{0000-0003-3347-7094}\,1}$\thanks{sharda@strw.leidenuniv.nl (PS)},
Shyam H. Menon$^{\orcidicon{0000-0001-5944-291X}\,2,3}$\thanks{shyam.menon@rutgers.edu (SHM)},
Roman Gerasimov$^{\orcidicon{0000-0003-0398-639X}\,4}$\thanks{rgerasim@nd.edu (RG)},
Volker Bromm$^{\orcidicon{0000-0003-0212-2979}\,5,6}$,
\newauthor
Blakesley Burkhart$^{\orcidicon{0000-0001-5817-5944}\,2,3}$,
Lionel Haemmerlé$^{\orcidicon{0000-0001-5614-5493}\,7}$,
Lisanne van Veenen$^{\orcidicon{0009-0005-8612-4059}\,1}$, and
Benjamin D. Wibking$^{\orcidicon{0000-0003-3175-2291}\,8}$
\\
$^{1}$Leiden Observatory, Leiden University, P.O. Box 9513, 2300 RA Leiden, The Netherlands\\
$^{2}$Department of Physics and Astronomy, Rutgers University, Piscataway, NJ 08854, USA\\
$^{3}$Center for Computational Astrophysics, Flatiron Institute, New York, NY 10010, USA\\
$^{4}$Department of Physics and Astronomy, University of Notre Dame, Notre Dame, IN 46556, USA\\
$^{5}$Department of Astronomy, University of Texas, Austin, TX 78712, USA\\
$^{6}$Weinberg Institute for Theoretical Physics, University of Texas, Austin, TX 78712, USA\\
$^{7}$Department of Astronomy, University of Geneva, CH-1290 Versoix, Switzerland\\
$^{8}$Department of Physics and Astronomy, Michigan State University, East Lansing, MI 48824, USA\\
}

\date{Accepted 2025 April 30. Received 2025 April 29; in original form 2025 January 20}

\pubyear{2024}

\begin{document}
\label{firstpage}
\pagerange{\pageref{firstpage}--\pageref{lastpage}}
\maketitle

\begin{abstract}
The masses of Population III stars are largely unconstrained since no simulations exist that take all relevant primordial star formation physics into account. We perform the first suite of radiation magnetohydrodynamics (RMHD) simulations of Population III star formation, with the POPSICLE project. Compared to control simulations that only include magnetic fields (MHD), protostellar ionizing and dissociating feedback, or neither, the RMHD simulation best resembles the MHD simulation during the earliest stages of collapse and star formation. In $5000\,\rm{yrs}$, the mass of the most massive star is $65\,\rm{M_{\odot}}$ in the RMHD simulation, compared to $120\,\rm{M_{\odot}}$ in simulations without magnetic fields. This difference arises because magnetic fields act against gravity, suppress mass transport, and reduce compressional heating. The maximum stellar mass of Population III stars is thus already limited by magnetic fields, even before accretion rates drop to allow significant protostellar radiative feedback. Following classical main sequence stellar evolution with MESA reveals that it is difficult to create Population III stars with masses larger than $600\,\rm{M_{\odot}}$ in typical dark matter minihaloes at $z \gtrsim 20$, with maximum stellar masses $\sim 100\,\rm{M_{\odot}}$ more likely due to expected negative feedback from both magnetic fields and stellar radiation. This work lays the first step in building a full physics-informed mass function of Population III stars.
\end{abstract}




\section{Introduction}
\label{s:intro}
Understanding the initial mass function (IMF) of Population~III (Pop~III) stars is of paramount importance, as evidenced by numerous theoretical works that examine the formation of these stars \citep[][and references therein]{Bromm:2013,2023ARA&A..61...65K}, as well as indirect observational evidence from metal poor stars \citep[e.g.,][]{2015ARA&A..53..631F,2019MNRAS.488L.109N,2021ApJ...915L..30S} and $z>10$ galaxies \citep[e.g.,][]{2023ApJS..265....5H,2023MNRAS.525.4832Y,2023arXiv230600953M}. The IMF is also crucial for determining whether Pop~III stars can be observed with JWST, or if their luminosity function can be differentiated from Pop~II stars in integrated light measurements \citep[][]{BKL2001,2002A&A...382...28S,2011ApJ...740...13Z,2024MNRAS.533.2727Z,2023MNRAS.525.5328T,fujimoto2025glimpseultrafaintsimeq105}. Equally important is to understand what sets the maximum possible mass (or, the upper mass cutoff of the IMF) of Pop~III stars \citep[e.g.,][]{2023MNRAS.522.3256C,Liu:2024,2024ApJ...962...49B}, which is an essential input for black hole seeding and supermassive star formation \citep[e.g.,][]{2018MNRAS.474.2757H,2022Natur.607...48L}. The upper mass cut off is also crucial to assess whether Pop~III stars of masses between $140-270\,\rm{M_{\odot}}$ existed, and could have exploded as pair-instability supernovae \citep[PISNe,][]{2010ApJ...724..341H,2017MNRAS.465..926D,2024ApJ...962L..26K}. Simulating the collapse of gas and formation of metal free stars in dark matter minihaloes provides a robust way to constrain the Population~III IMF and its upper mass cut off. 

However, radiation-magnetohydrodynamics simulations in the era of Pop~III star formation remain largely absent. Both magnetic fields and protostellar radiation feedback are critical ingredients that influence (massive) star formation across all metallicities \citep[e.g.,][]{2018ApJ...861...68T,2021MNRAS.508.4175C,2023arXiv231213339C,2022MNRAS.509.1959S}. Therefore, conclusions from prior numerical work aimed at deriving the masses of Pop~III stars are likely subject to major uncertainties since they either exclude magnetic fields \citep[e.g.,][]{2011Sci...334.1250H,2016ApJ...824..119H,2020ApJ...892L..14S,2022MNRAS.512..116J} or radiation feedback \citep[e.g.,][]{2012ApJ...745..154T,2020MNRAS.497..336S,2022MNRAS.516.2223P,2022MNRAS.516.3130S,2023MNRAS.519.3076S}. In this work, we use the POPSICLE project (Sharda et al. in prep.) to extend the recent RMHD simulations of Pop~III star formation by \citet{2024arXiv240518265S} to include far-UV (FUV) molecule-dissociating radiation feedback in addition to extreme-UV (EUV) ionizing feedback. Our aim in this Letter is to show that magnetic fields significantly limit the mass growth of massive Pop~III stars, \textit{even before} radiative feedback becomes dominant. We arrange the remainder of the paper as follows: \autoref{s:setup} summarizes the setup we use to develop and run the simulations, and \autoref{s:results} describes the results. In \autoref{s:mesa}, we look at the main sequence evolution of simulated stars using Modules for Experiments in Stellar Astrophysics \citep[MESA,][]{2011ApJS..192....3P,MESA_2,MESA_3,MESA_4,2019ApJS..243...10P}. Finally, we summarize in \autoref{s:conclusions}. 

\section{POPSICLE Simulations}
\label{s:setup}
We briefly describe the POPSICLE\footnote{\textbf{POP}ulation II/III \textbf{S}imulations \textbf{I}ncluding \textbf{C}hemistry, \textbf{L}uminosity, and \textbf{E}lectromagnetism.} project setup employed in this work, and point the reader to \citet{2024arXiv240518265S} for details of the numerical implementation. We use a custom version of the adaptive mesh refinement (AMR) code FLASH \citep{2000ApJS..131..273F,2008ASPC..385..145D} which employs an approximate Riemann solver for magnetohydrodynamics \citep{2009JCoPh.228.8609W,2011JCoPh.230.3331W}. We include non-equilibrium primordial chemistry (with deuterium) from the KROME astrochemistry package \citep{2014MNRAS.439.2386G}. We use the VETTAM radiation hydrodynamics scheme, which uses a non-local variable Eddington tensor (VET) closure obtained with a ray-trace solve to close the radiation moment equations \citep{2022MNRAS.512..401M}. We use the \texttt{GENEVA} Pop~III protostellar model grid to evolve radiative properties of the protostars as a function of their mass and accretion rates \citep{2016A&A...585A..65H,2018MNRAS.474.2757H}. The protostars are represented by sink particles in the simulation, following the criteria described in \citet{2010ApJ...713..269F,2011IAUS..270..425F}.

Population III protostars contract and produce significant radiation once accretion rates drop below $0.01\,\rm{M_{\odot}\,yr^{-1}}$ \citep{op2001,op2003}. We consider the ionization of H and H$_2$ due to extreme-UV (EUV) photons released from the protostar(s) with energies upwards of $13.6\,\rm{eV}$ (upwards of $15.2\,\rm{eV}$ for H$_2$). The key physics we add to the simulations by \citet{2024arXiv240518265S} is the dissociation of H$_2$ by FUV photons in the Lyman-Werner (LW) band, between $11.2 - 13.6\,\rm{eV}$, which can significantly affect the thermodynamic state of dense gas in the vicinity of the stars. Importantly, we consider both self-shielding of H$_2$ as well as cross shielding by H \citep[][]{2011MNRAS.412.2603W}; the latter has been ignored in previous works. We adopt the fitting functions for self- and cross-shielding of H$_2$ from \citet{2011MNRAS.418..838W}. We do not invoke an external LW background in addition to in-situ LW radiation from the protostars\footnote{This approximation is supported by the build-up of local opacity to external LW radiation due to relic \ion{H}{II} regions in the early intergalactic medium \citep{Johnson2007}.}. 

We keep the initial conditions identical to \citet{2024arXiv240518265S}: the initial cloud mass and radius are $1000\,\rm{M_{\odot}}$ and $1\,\rm{pc}$, respectively. We also include initially trans-sonic turbulence within the cloud, and impose solid body rotation with rotational energy equal to 3 per cent of the gravitational energy. We refine using 64 cells per Jeans length, significantly higher than published simulations that include radiation feedback even at Solar metallicity. With 10 levels of grid refinement based on the Jeans length, the maximum spatial resolution we achieve is $\Delta x = 7.5\,\rm{au}$. Our sink particle density threshold is $10^{13}\,\rm{cm^{-3}}$. Based on earlier results that show how even initially weak magnetic fields are quickly amplified to saturation due to the small-scale dynamo \citep[e.g.,][]{2012ApJ...754...99S,2015PhRvE..92b3010S,2012ApJ...745..154T,2020MNRAS.497..336S,2021MNRAS.503.2014S}, we set our initial magnetic field strength to be $28\,\rm{\micro G}$, equivalent to 10 per cent of the initial turbulent kinetic energy, appropriate for parsec scales in the presence of trans-sonic turbulence and low magnetic Prandtl numbers \citep[for details, see][their fig.~4]{2020MNRAS.497..336S,Hirano:2018}. The power spectrum of the trans-sonic turbulence we initially drive follows $P_{\rm{v}} \propto k^{-1.8}$, in between the Kolmogorov \citep[$k^{-5/3}$,][]{1941DoSSR..32...16K} and Burgers \citep[$k^{-2}$,][]{BURGERS1948171} scalings. The turbulence consists of a mixture of solenoidal and compressive modes \citep[][]{2010A&A...512A..81F}. The initial magnetic field is completely random with no preferred orientation, as expected for a small-scale turbulent dynamo. 

We compare to control simulations without magnetic fields or radiation hydrodynamics (HD), and with only magnetic fields (MHD) from \citet{2024arXiv240518265S}. The turbulent realization we select from \citet{2024arXiv240518265S} to control for stochasticity is the one that produces only one star in the HD case, since this case is straightforward to analyze. In addition to this, we run a control simulation with only radiation feedback for the same realization (including both EUV and FUV feedback, termed RHD). These control simulations are instructive as they can help distinguish between the impact of magnetic fields and radiation feedback on Pop~III star formation.

\begin{figure}
\includegraphics[width=\columnwidth]{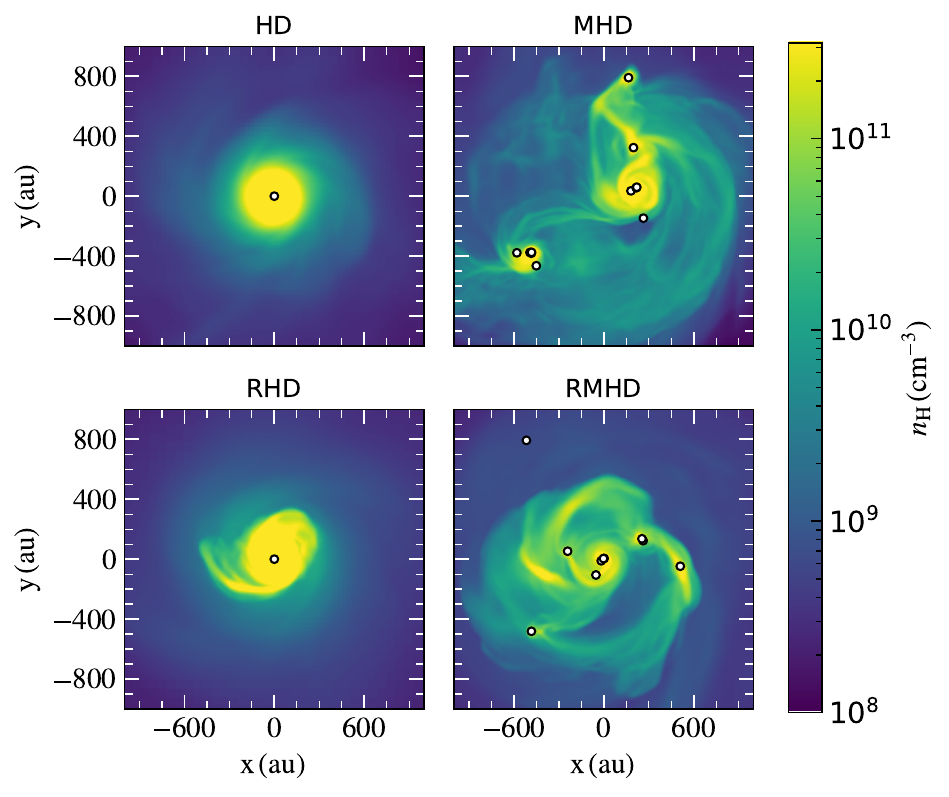}
\caption{Face-on density-weighted projections of the gas number density along the $\hat{z}$ axis, at the end of the simulations ($5000\,\rm{yr}$ post the formation of the first star). The four panels correspond to the runs with hydrodynamics (HD), magneto-hydrodynamics (MHD), radiation-hydrodynamics including ionizing and dissociating radiation feedback (RHD), and radiation-magnetohydrodynamics (RMHD). White dots represent the position(s) of sink particle(s) that form, used as a proxy for Pop~III stars.}
\label{fig:proj_numdens}
\end{figure}

\begin{figure}
\includegraphics[width=\columnwidth]{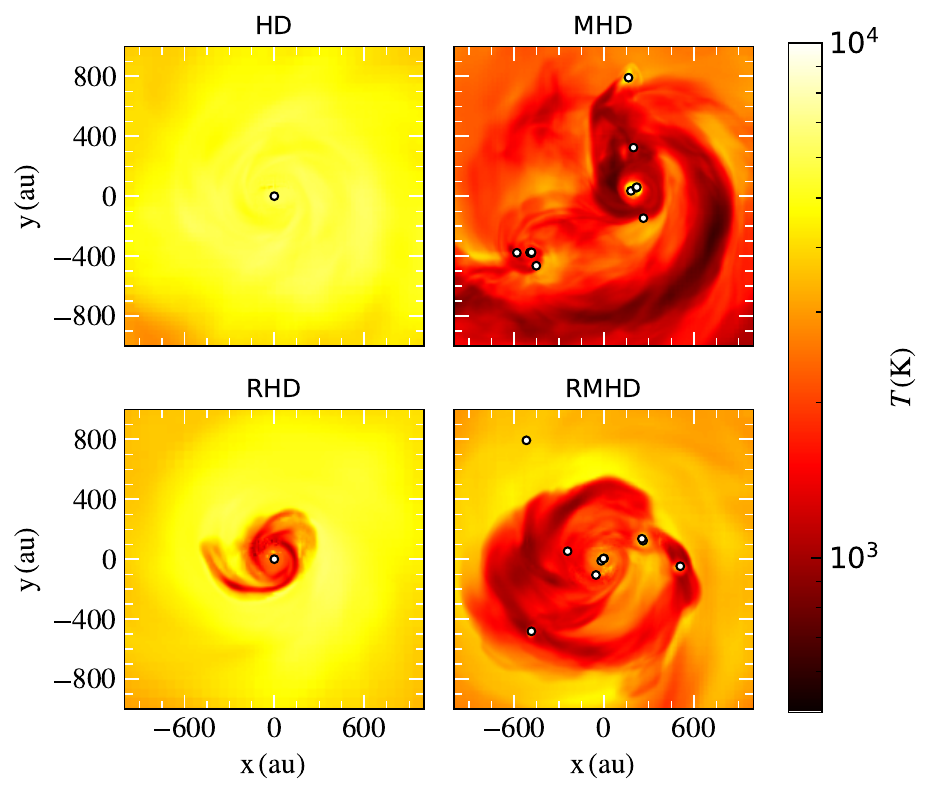}
\caption{Same as \autoref{fig:proj_numdens} but for the density-weighted gas temperature at the end of the four simulations.}
\label{fig:proj_temp}
\end{figure}

\begin{figure}
\includegraphics[width=\columnwidth]{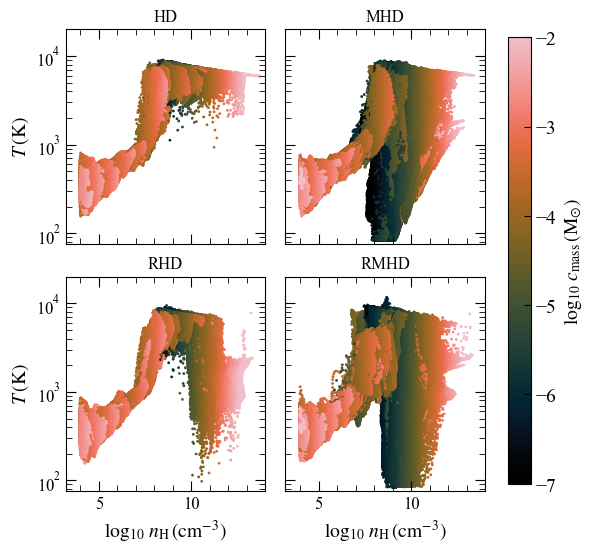}
\caption{Gas temperature ($T$) and density ($n_{\rm{H}}$) phase diagrams for all the cells within the cloud at the end of the simulations ($5000\,\rm{yr}$ post the formation of the first star), color-coded by the cell mass, $c_{\rm{mass}}$.}
\label{fig:phase}
\end{figure}

\section{Results}
\label{s:results}
We evolve the simulations until 5000~yr post the formation of the first star. We find that both runs including magnetic fields show fragmentation, leading to the formation of Pop~III star clusters. This means that the evolution of the most massive Pop~III star in the MHD and RMHD runs is influenced by companion stars. However, with only one turbulent realization, we lack the statistics to quantify the impact of magnetic fields on fragmentation in primordial clouds. In fact, other turbulent realizations presented in \citet{2024arXiv240518265S} fragment even in the HD case (see also, discussions in \citealt{2020MNRAS.494.1871W} and \citealt{2020MNRAS.497..336S} on stochastic fragmentation due to turbulence).

\autoref{fig:proj_numdens} plots the density-weighted projections of the number density of the gas at the end of the simulations. The projection window is $0.01\,\rm{pc}$ wide. It is centered on the isolated star in the HD and RHD runs, and uses the center of mass of all stars in the MHD and RMHD runs. We plot the density-weighted projections of the gas temperature for the four simulations in \autoref{fig:proj_temp}, and phase diagrams for the entire cloud in \autoref{fig:phase}. We see from \autoref{fig:proj_numdens} and \autoref{fig:proj_temp} that the central star in the HD and RHD runs is fed by a well-defined, thermal pressure-supported accretion disc that remains stable against fragmentation for the entirety of the simulation. The high temperatures in these runs is a result of dissociation of H$_2$ by shocks, a process that is only resolved in simulations with sufficiently high Jeans resolution \citep[e.g.,][]{2012ApJ...745..154T,2021MNRAS.503.2014S,2024arXiv240518265S}. The key difference between the HD and RHD runs is the presence of cooler, H$_2$-dominated gas near the star in the latter, where H$_2$ dissociation is prevented due to radiation pressure slowing down accretion shocks (see also, \citealt{2024arXiv240518265S}). Given the relatively high accretion rates, the star is unable to contract and produce significant ionizing (EUV) photons that can ionize H or H$_2$ \citep[e.g.,][]{op2001,2009ApJ...691..823H,2018MNRAS.474.2757H}. In contrast to HD and RHD simulations, the gas is cooler in MHD and RMHD runs across the entire central envelope. Magnetic fields suppress gravitational collapse, thereby also reducing the rate of compressional heating. Reduced heating allows the gas to remain molecular, and H$_2$ cooling further ensures gas temperatures remain low. 

These figures show, for the first time, the simultaneous effects of magnetic fields, and ionizing as well as dissociating radiation feedback during the evolution of primordial clouds. Note that we do not include the effects of radiative heating caused by accretion luminosity, which could lead to higher gas temperatures close to the protostars, thereby reducing the accretion rates and limiting protostellar mass growth \citep[e.g.,][]{2011MNRAS.414.3633S,2020MNRAS.494.1871W}. However, it cannot hinder accretion for long time periods since low accretion rates in turn lead to lower accretion heating. Further, accretion luminosity heating only becomes significant when the opacity of (primordial) gas is large, which occurs for $T > 5000\,\rm{K}$ at the densities we resolve \citep[][Table E3]{2005MNRAS.358..614M}. Given that magnetic fields lead to lower average gas temperatures, we expect accretion luminosity to be less important in the MHD and RMHD simulations as compared to HD and RHD simulations.

\begin{figure}
\includegraphics[width=\columnwidth]{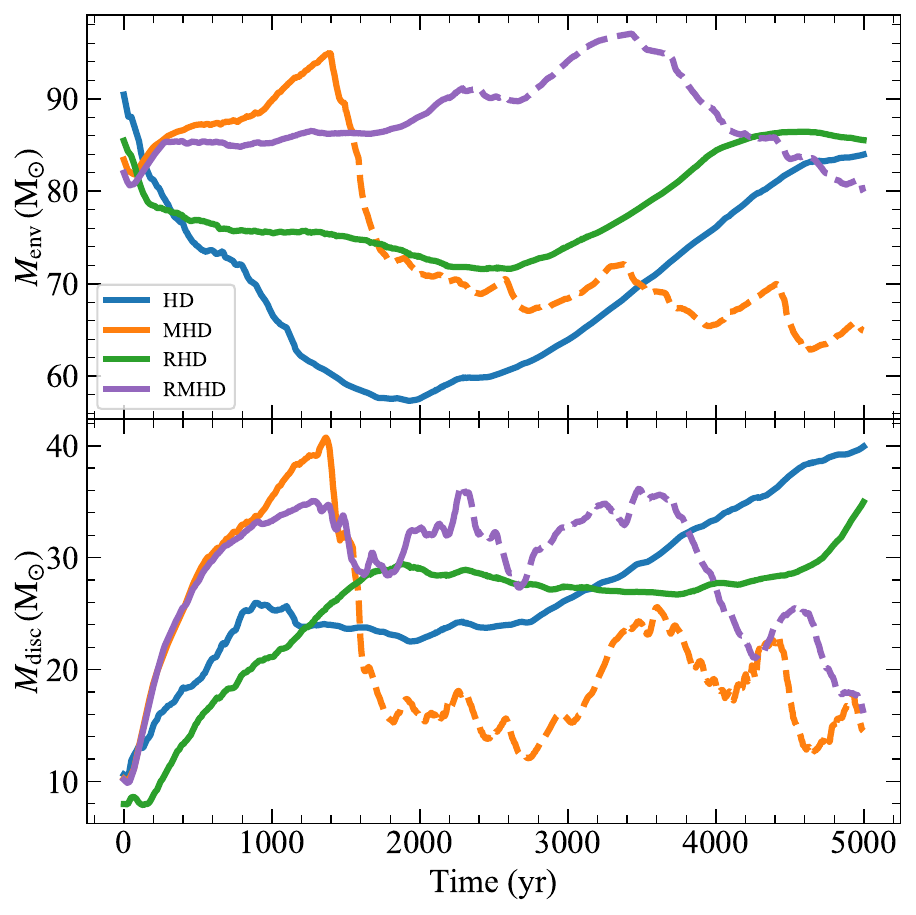}
\caption{\textit{Top panel:} evolution of mass enclosed within an envelope of radius $0.01\,\rm{pc}$, centered at the location of the most massive star in each simulation. The MHD and RMHD runs fragment $1500\,\rm{yr}$ and $2235\,\rm{yr}$ after the formation of the first star, which is demarcated by the onset of dashed orange and purple curves, respectively. \textit{Bottom panel:} mass enclosed within a disc of radius $500\,\rm{au}$ and height $50\,\rm{au}$ (from the midplane) around the most massive star. The mass reservoir that can be accreted onto the central star in the MHD and RMHD runs eventually decreases as magnetic fields suppress gravitational collapse.}
\label{fig:core}
\end{figure}

We now turn to look at the time evolution of gas properties that dictate the mass growth of the protostars. The two panels of \autoref{fig:core} plot the evolution of the mass enclosed within an envelope surrounding the accretion disc-star system, and the accretion disc itself in the four simulations. Following \citet{2021MNRAS.503.2014S} and \citet{2024arXiv240518265S}, we define the envelope to be a $0.01\,\rm{pc}$ spherical region centered on the most massive star. Similarly, we define the accretion disc as a cylindrical region of radius $500\,\rm{au}$ and $50\,\rm{au}$ in height (from the midplane). The dashed curves in the MHD and RMHD runs mark the onset of fragmentation, emphasizing that the most massive star does not evolve in isolation thereafter.

\begin{figure}
\includegraphics[width=\columnwidth]{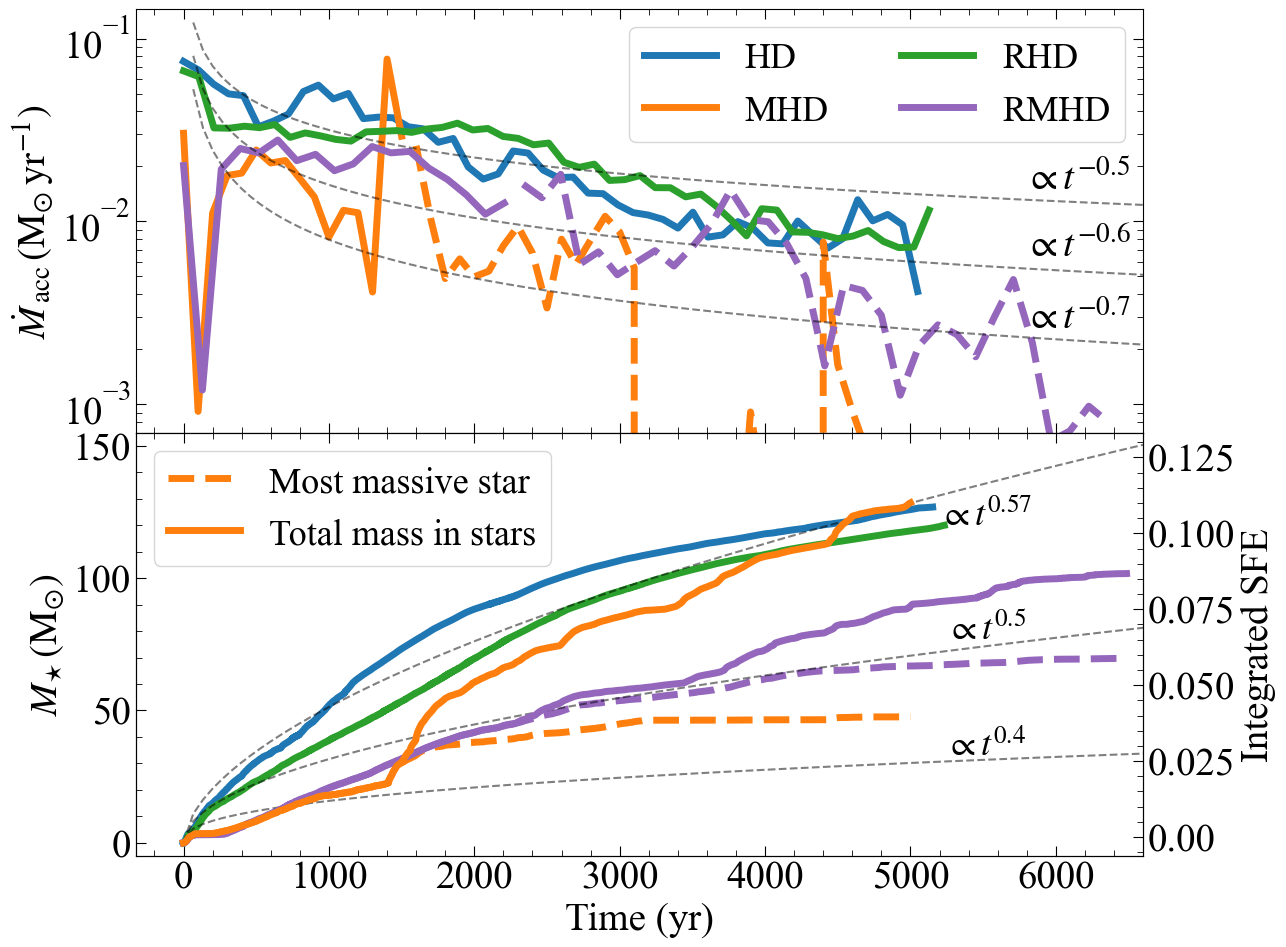}
\caption{\textit{Top panel:} gas accretion rate onto the most massive star as a function of time in the four simulations, averaged over $100\,\rm{yr}$ intervals. As in \autoref{fig:core}, the transition to dashed curves in the MHD and RMHD runs reflect the onset of fragmentation within the collapsing core. Background grey curves depict example accretion rate profiles of the form $\dot M_{\rm{acc}} \propto t^{\alpha}$ with $\alpha = -0.5, -0.6, -0.7$. \textit{Bottom panel:} cumulative mass growth of all the stars (solid) and that of the most massive star (dashed) in the four simulations. Dashed grey curves demarcate example trends of the form $M_{\star} \propto t^{\gamma}$ with $\gamma = 0.57, 0.5, 0.47$ and $0.4$. Lower accretion rates in the MHD and RMHD runs lead to lower maximum stellar masses (by a factor $\approx 2$ ) as compared to the HD and RHD runs. The integrated star formation efficiency (SFE) is lower by 30 per cent in the RMHD run compared to all other runs.}
\label{fig:mass}
\end{figure}

We see from \autoref{fig:core} that the amount of mass reaching the envelope initially increases in the MHD and RMHD runs, but then turns over and starts to decline. Despite the envelope and the disc containing larger masses early on in the MHD and RMHD runs, the corresponding accretion rates onto the protostars are lower (see \autoref{fig:mass}). Close to the protostar, the magnetic field is sub-Alfvénic (plasma $\beta < 1$), significantly inhibiting mass transport. In contrast, the rate of mass transfer from the envelope to the accretion disc is initially quite fast in the HD and RHD runs, resulting in a decline of mass in the envelope and buildup of mass in the disc. This mass is consequently accreted by the star at a high rate. Fragmentation events correlate with a large buildup of mass in the accretion disc such that the ratio of mass in the disc to the protostellar mass increases far beyond unity. The presence of multiple stars reduces the mass available within the disc of the primary star in the MHD and RMHD runs, whereas the disc mass in the HD and RHD runs continues to build up.

The top panel of \autoref{fig:mass} shows that the accretion rates (of the most massive star) are of the order of $0.03\,\rm{M_{\odot}}\,yr^{-1}$ in the first $1000\,\rm{yr}$, and decline as time progresses. The accretion rates in the MHD and RMHD runs remain systematically lower than the HD and RHD runs for the most part, initially due to strong magnetic fields inhibiting accretion, and later due to fragmentation. The key impact of lower accretion rates in the MHD and RMHD runs is that the most massive stars only build up half as much mass as the HD and RHD runs within the same time period. We show this in the bottom panel of \autoref{fig:mass}. The mass of the (isolated) star at the end of the simulations in the HD and RHD runs is $127\,\rm{M_{\odot}}$ and $120\,\rm{M_{\odot}}$, respectively. On the other hand, the mass of the most massive star in the MHD and RMHD runs is $48\,\rm{M_{\odot}}$ and $67\,\rm{M_{\odot}}$, respectively. However, the integrated star formation efficiency (SFE; defined as the cumulative mass of all stars normalized by the initial cloud mass; solid orange and purple curves in the bottom panel of \autoref{fig:mass}) is lower by 30 per cent in the RMHD run as compared to all the control runs. 

There is a subtle but important difference between magnetic fields suppressing mass transport, and fragmentation-induced starvation. The former turns on earlier, and affects both the mass of the primary star and the integrated SFE (dashed and solid lines in the bottom panel of \autoref{fig:mass}), whereas the latter emerges later, and affects the mass of the primary star mass without necessarily affecting the integrated SFE. Thus, the `mass-limiting' effects of the runs including magnetic fields are not simply due to fragmentation-induced starvation. The slower mass growth and consequently lower star formation efficiency in the presence of magnetic fields has also been observed in simulations of Population I massive star formation, although the differences arise much later on, and are smaller in magnitude \citep[e.g.,][]{2020AJ....160...78R,2021ApJ...911..128K}. Our findings are also in qualitative agreement with \cite{2024arXiv240518265S}, where the authors simulated Pop~III star formation with magnetic fields and only ionizing feedback. 

\begin{figure*}
\includegraphics[width=\textwidth]{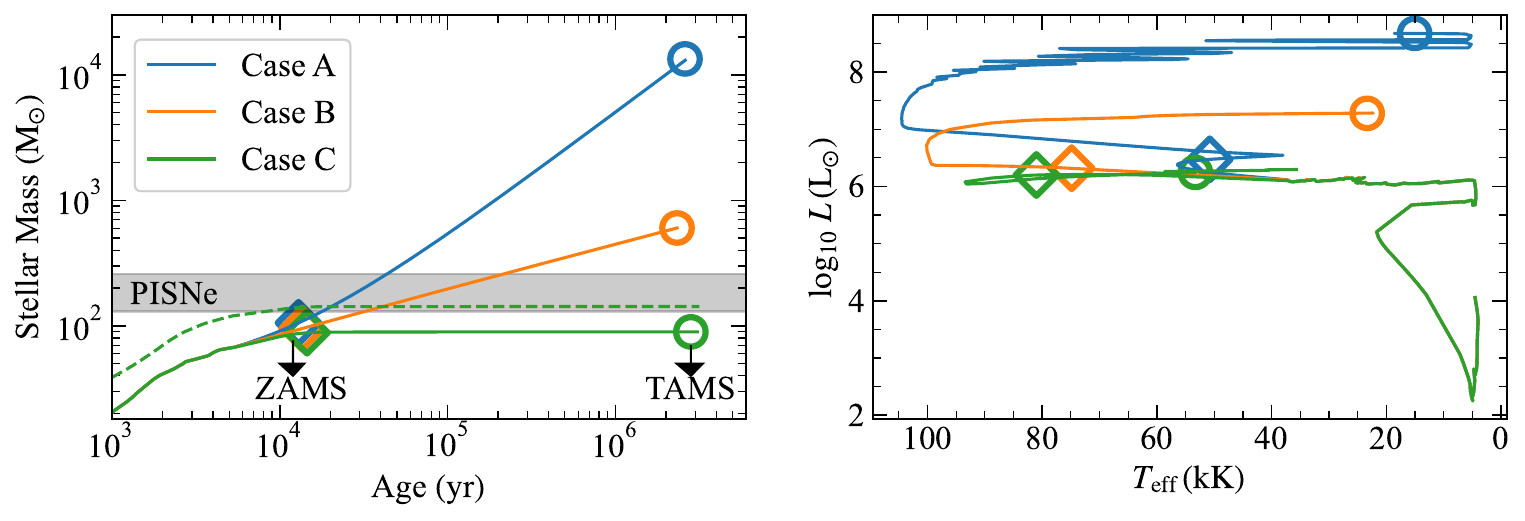}
\caption{\textit{Left panel:} stellar mass as a function of age of the most massive star in the RMHD run, extrapolated beyond the period simulated using three distinct accretion histories. Cases A, B, and C represent progressively steeper (and likely more realistic) decline in accretion rates over time (see \autoref{s:mesa} for details). Diamonds and circles mark the age at which the star reaches zero age main sequence (ZAMS) and terminal age main sequence (TAMS) respectively in the MESA calculations. The TAMS masses suggest all the three cases lead to the star ending its life as a black hole. Grey-shaded region denotes stars that will explode as pair-instability supernovae (PISNe). Dashed green curve marks the growth of the star in the RHD simulation extrapolated using Case C. \textit{Right panel:} HR diagram of the star in the RMHD simulation as it evolves on the main sequence under the three cases presented in the left panel.}
\label{fig:final}
\end{figure*}

\section{Evolution on the main sequence}
\label{s:mesa}
In this section, we explore the implications for the long-term mass growth of Pop~III stars and the upper mass cut-off of the Pop~III IMF. Fully simulating this process is beyond the scope of this work. Nonetheless, we can use the information available from our simulations to predict the timescales to reach the zeroage main sequence (ZAMS), as well as the ultimate fate as these stars reach the terminal age main sequence (TAMS). For this purpose, we run stellar structure evolution calculations using MESA, version 23.05.1 \citep{2011ApJS..192....3P,MESA_2,MESA_3,MESA_4,2019ApJS..243...10P}. \aref{s:app_mesa} lists the details of the modeling.

We assume no accretion in our MESA models for the first $\sim 25$ years, corresponding to the amount of time the most massive star in the RMHD simulation takes to reach the adopted initial stellar mass in MESA ($1\,\mathrm{M}_\odot$). After that, we use the RMHD accretion history of said most massive star, and consider three increasingly realistic scenarios for the evolution beyond the final time of our simulations ($5000\,\rm{yr}$).
In Case A (\autoref{fig:final}), we assume the star continues to accrete at a constant rate of $0.005\,\rm{M_{\odot}\,yr^{-1}}$, corresponding to the average accretion rate over the final $1000\,\rm{yr}$ of the simulation. Such an accretion history for the entirety of the stellar lifetime is rather unrealistic, but we include it to provide a baseline for comparison. In Case B, we assume the average accretion rate beyond $5000\,\rm{yr}$ follows the trend observed in the simulations, decreasing with time as $t^{-0.65}$. In Case C, we empirically include the effects of radiation feedback that can halt accretion at late times. To do so, we define an accretion rate that goes as $e^{-t^2}$ between $5000 \leq t \leq 2\times10^4\,\rm{yr}$ and linearly declines as $t^{-3}$ beyond $2\times 10^4\,\rm{yr}$, mimicking the trend seen in radiation hydrodynamics simulations of \citet[][Fig. 3]{2011Sci...334.1250H}. 

We plot the resulting mass growth from the three cases as blue, orange and green curves in \autoref{fig:final}, respectively. Despite different accretion histories, the star in all the three cases reaches ZAMS around $13000\,\rm{yr}$ (marked by diamonds in \autoref{fig:final}) with stellar mass between $90 - 100\,\rm{M_{\odot}}$. However, the main sequence evolution is significantly different, and the TAMS mass spans a large range. The stars are expected to turn into black holes soon after TAMS (denoted by circles in \autoref{fig:final}). The star in Case A oscillates between two branches, leading to wiggles in the HR diagram, because its evolution is sensitive to atmospheric parameters \citep[][]{2023MNRAS.521..463H}. This phenomenon only occurs for Case A because the accretion rate is close to the critical value of $0.01\,\rm{M_{\odot}}\,yr^{-1}$ which separates the blue and red branches of supermassive stellar evolution \citep[][]{2009ApJ...691..823H,2018MNRAS.474.2757H}. The period we simulate with different physics, although short, matters for the final fate of the star. To show this, we use Case C to extrapolate the growth of the star in the RHD simulation (dashed green curve in the left panel of \autoref{fig:final}). Due to differences in accretion histories in the first $5000\,\rm{yr}$, this star ends into the regime of PISNe, rather than ending its life as a black hole. This is unlikely to occur in reality since magnetic fields are excluded in the RHD simulation, but it shows that omitting key physical processes even during the early stages of evolution can profoundly impact the ultimate fate of Pop~III stars.

Given the high stellar effective temperatures and luminosities on the main sequence (right panel of \autoref{fig:final}), radiation feedback combined with competitive accretion due to fragmentation will potentially limit the final mass to $80\,\rm{M_{\odot}} \leq M_{\star} \lesssim 600\,\rm{M_{\odot}}$, making Case A highly unlikely. The magnetic field strength is expected to remain significant beyond the period we simulate here due to a mixture of turbulent and mean field dynamos \citep[][]{2019arXiv191107898L,2021MNRAS.503.2014S}, and it can provide additional negative feedback by suppressing accretion onto the star, or generating outflows \citep{2013MNRAS.435.3283M}. Together with the effects of heating due to accretion luminosity we discuss in \autoref{s:results}, this will enhance feedback effects and further limit the maximum stellar mass, such that the actual stellar mass is closer to $80\,\rm{M_{\odot}}$ than $600\,\rm{M_{\odot}}$.

\section{Summary}
\label{s:conclusions}
We use the POPSICLE simulations suite to perform radiation-magnetohydrodynamics (RMHD) simulations of Population III star formation. We simultaneously include non-equilibrium primordial chemistry, turbulent magnetic fields, ionizing and dissociating stellar feedback, following the evolution $5000\,\rm{yr}$ post the formation of the first star. We also carry out control simulations where we use identical initial conditions but only include magnetic fields (MHD), protostellar radiation feedback (RHD), or exclude both (HD).

We find that, during the earliest stages we simulate, magnetic fields suppress gravitational collapse, leading to less compressional heating and inefficient mass transport from the cloud to the protostar. As a result, the gas temperature in the MHD and RMHD runs are lower than the HD and RHD runs, which makes the gas more susceptible to fragmentation. Both the runs including magnetic fields (MHD and RMHD) fragment, however, with only one turbulent realization, we lack the statistics to make quantitative conclusions on how fragmentation occurs in the presence of both magnetic fields and radiation feedback \citep[e.g.,][]{2020MNRAS.494.1871W,2020MNRAS.497..336S}. The combined effect of suppression of mass transport to the star and fragmentation-induced starvation is that the mass of the most massive star in the RMHD run is $65\,\rm{M_{\odot}}$, factor of two lower than that in the HD and RHD runs. 

Radiation feedback has long been proposed as the primary mechanism that halts the growth of Pop~III stars and sets the upper mass cutoff of the Pop~III IMF \citep[][]{2011Sci...334.1250H,2016ApJ...824..119H,Stacy:2012,Stacy:2016}. Here, we show that magnetic fields can also limit mass growth, even before accretion rates drop to facilitate significant ionizing or dissociating feedback \citep[][]{2018MNRAS.474.2757H}. Because magnetic fields also slow down accretion onto protostars, they can induce strong radiation feedback earlier than expected. If a mean field dynamo operates and is sustained for long periods of time, magnetic fields will also enable launching protostellar outflows, further reducing the maximum possible mass of Pop~III stars \citep[][]{2013MNRAS.435.3283M,2019arXiv191107898L,2021MNRAS.503.2014S}. 

Stellar structure modeling with MESA using the accretion history from the RMHD simulation, extrapolated considering a range of scenarios for subsequent mass growth, shows that the $65\,\rm{M_{\odot}}$ star reaches ZAMS early on at an age of $13,000\,\rm{\,yr}$, and likely continues to accrete on the main sequence. The star then evolves off the main sequence in about $2.5\,\rm{Myr}$. The initial evolutionary phase has implications both for the final mass and fate (supernova versus black hole) of the star. The final mass spans a large range based on possible accretion histories, from $80\,\rm{M_{\odot}}$ to $600\,\rm{M_{\odot}}$, although combined effects of radiation feedback, magnetic fields and fragmentation will render the actual mass closer to the lower bound. This work lays the foundation for constructing a Pop~III IMF with all relevant star formation physics. Future work will involve simulating multiple realizations to build a mass distribution, following the accretion history for longer time periods to quantify feedback at late times, and exploring black hole seeding from Pop~III stars.

\section*{Acknowledgements}
We thank Zoltan Haiman, Simon Glover, Mark Krumholz, Mariska Kriek and Matthieu Schaller for useful discussions, and the anonymous referee for their feedback that helped improve this work. We also thank SURF for the support in using the Dutch National Supercomputer Snellius. PS is supported by the Leiden University Oort Fellowship and the International Astronomical Union -- Gruber Foundation (TGF) Fellowship. SHM and BB acknowledge support through NASA grant No. 80NSSC20K0500 and National Science Foundation (NSF) grant AST-2009679, and the CCA at the Flatiron Institute. RG is supported by the Society of Science Fellowship provided by the University of Notre Dame. BB is grateful for generous support by the David and Lucile Packard Foundation and Alfred P. Sloan Foundation. BDW acknowledges support from NSF grant AAG-2106575. The simulations and data analyses presented in this work used high-performance computing resources provided by the Dutch National Supercomputing Facility SURF via project grant EINF-8292 on Snellius, as well as high-performance computing resources provided by the Centre for Computational Astrophysics (CCA) at the Flatiron Institute, and the Australian National Computational Infrastructure (NCI) through project \texttt{jh2} in the framework of the National Computational Merit Allocation Scheme and the Australian National University (ANU) Allocation Scheme. This work was performed in part at Aspen Center for Physics, which is supported by National Science Foundation grant PHY-2210452. We acknowledge using the following softwares: FLASH \citep{2000ApJS..131..273F,Dubey_2013}, \texttt{VETTAM} \citep{2022MNRAS.512..401M}, \texttt{KROME} \citep{2014MNRAS.439.2386G}, \texttt{petsc} \citep{petsc-efficient,petscsf2022}, Astropy \citep{2013A&A...558A..33A,2018AJ....156..123A,2022ApJ...935..167A}, Numpy \citep{oliphant2006guide,2020arXiv200610256H}, Scipy \citep{2020NatMe..17..261V}, Matplotlib \citep{Hunter:2007}, \texttt{yt} \citep{2011ApJS..192....9T}, \texttt{cmasher} \citep{2020JOSS....5.2004V}. This research has made extensive use of NASA's Astrophysics Data System Bibliographic Services (ADS).

\section*{Data Availability}
Movies associated with this article are available as supplementary (online only) material. To access the suite of simulations or properties of the stars, please contact the authors.



\bibliographystyle{mnras}
\bibliography{references} 


\appendix
\section{Details of MESA modeling}
\label{s:app_mesa}
In general, MESA simulations cannot account for the early phases of stellar evolution, as the simulated object is required to be in approximate hydrostatic equilibrium (up to a set of implemented dynamical corrections) and obey the equations of stellar structure. For this reason, it is necessary to choose a non-zero initial mass and the initial structure of the star at the zero age of the simulation. We initialize all MESA simulations with a metal-free composition ($X=0.75$, $Y=0.25$, $Z=0$), initial mass $1\ \mathrm{M}_\odot$, and initial central temperature $T_c=61500\ \mathrm{K}$. For the accretion rate predicted by the RMHD simulation, the chosen values approximately correspond to the lowest initial mass and $T_c$, for which the model can converge. In order to capture as much of the early stellar evolution as possible, it is desired to choose the smallest possible values of initial mass and $T_c$, as they likely represent the earliest phases of the pre-main sequence contraction. MESA uses an adaptive time step that depends on the rate of evolution; however, we found the default implementation of the adaptive algorithm too generous to capture the subtleties of the pre-main sequence evolution with rapid accretion. In all three cases, we average the accretion history provided to MESA in $10\ \mathrm{yr}$ steps to suppress discontinuities in the evolutionary tracks.

We define the ZAMS point as the earliest age, at which $0.1\%$ of the hydrogen content at the center of the star has been used up by nuclear fusion. The TAMS is typically defined by the complete exhaustion of hydrogen in the core, which is subsequently followed by the onset of the red giant branch, characterized by ignition of nuclear fusion away from the center of the star \citep{RGB_definition}. However, in a massive Pop~III star, the latter process may begin before the core hydrogen reservoir is fully exhausted, making the TAMS point less distinct. The carbon-nitrogen-oxygen (CNO) cycle serves as the primary energy source in the cores of massive Population~III stars \citep{larkin2023} even for initially metal-free compositions. On the other hand, a rapid increase in the energy production rate by the proton-proton chain may be used as an early signature of nuclear burning outside the core, where the conditions are less favorable for the CNO cycle. We therefore define TAMS as the earliest age, at which 1.) the energy production rate due to the proton-proton chain is increasing, and 2.) the central hydrogen fraction has dropped below 5 per cent. The first condition is necessary as the energy production of the proton-proton chain is also expected to increase during the pre-main sequence evolution.

\bsp	
\label{lastpage}
\end{document}